# Opportunities and challenges in the use of heavily doped polycrystalline silicon as a thermoelectric material. An experimental study.

Dario Narducci[#1], Ekaterina Selezneva[#2], Gianfranco Cerofolini[#3], Elisabetta Romano[#4],
Rita Tonini[*5], and Giampiero Ottaviani[*6]

[#]CNISM and Department of Materials Science, University of Milano-Bicocca, via R. Cozzi 53, 20125 Milano, Italy
[1]dario.narducci@unimib.it
[2]ekaterina.selezneva@mater.unimib.it
[3]gianfranco.cerofolini@mater.unimib.it
[4]elisabetta.romano@mater.unimib.it

[*]Department of Physics, University of Modena and Reggio Emilia, Via Campi 213, 41100 Modena, Italy
[5]rita.tonini@unimore.it
[6]giampiero.ottaviani@unimore.it

*Abstract*—Large-volume deployment of Si-based Seebeck generators can be foreseen only if polycrystalline rather than single crystalline materials can be actually used. The aim of this study was therefore to verify whether polycrystalline Si films deposited on top of a $SiO_2$ insulating layer can develop interesting thermoelectric power factors. We prepared 450-nm thick heavily boron doped polysilicon layers, setting the initial boron content in the film to be in excess of the boron solubility in polycrystalline silicon at 1000 °C. Isochronal thermal annealings were then used to modify the $B_{Si}$ content by precipitation. Quite unexpectedly, a concurrent increase of the thermoelectric power and of the conductivity was observed for heat treatments at temperatures above 800 °C. Upon annealing at 1000 °C we found a power factor $P$ of 13 mW $K^{-2}$ $m^{-1}$, more than three times higher than previously reported $P$ for Si nanowires. These findings could be explained observing that degenerate polysilicon displays a remarkable enhancement of its Seebeck coefficient as an effect of the large amount of boron it can dissolve. Band gap narrowing and band tailing modify the density of states around the Fermi energy leading to a dramatic improvement of its log-derivative in the Mott equation. These results apparently point out an interesting direction for the development of Seebeck and Peltier devices sharing low cost and relatively high efficiency.

## I. INTRODUCTION

Search for high efficiency thermoelectric materials has undergone a recent acceleration, also possibly in connection with the appearance of two papers [1], [2] showing the possibility, opened by nanostructuring, of using a notoriously poorly performing material as silicon to obtain devices with interesting figures of merit. Actually, while materials such as tellurides are characterized by relatively high efficiency, it is hard to believe, as recently recalled, a wide spread of thermoelectric converters based upon uncommon, high-cost materials [3]. Instead, applications based upon silicon might qualify for distributed energy recover.

Silicon nanowires and other dimensionally constrained systems were demonstrated to tackle the issue of increasing the material figure of merit $zT$ by decreasing the material thermal conductivity $\kappa$ while not lowering either the Seebeck coefficient $\alpha$ or the electrical conductivity $\sigma$, so that $zT$ (= $\alpha^2 \sigma T / \kappa$) could reach values close to 1 at $T$ = 300 K. On the other hand, the power factor $P = \alpha^2 \sigma$ is intrinsically limited in diluted semiconductors by the functional dependency of $\alpha$ and $\sigma$ upon the carrier density [4]. In such a density range, neglecting the marginal mobility dependence upon ionized impurities and referring e.g. to p-type materials, the thermopower scales with the hole density $p$ as $|\alpha| \propto \ln(1/p)$ and $\sigma \propto p$. Actually, in single-crystal Si and based upon early investigations [5], [6] it has been computed [7], [8] that the optimal carrier density for $zT$ is in the range $10^{18}$–$10^{19}$ $cm^{-3}$ [9]. However, interesting possibilities of overcoming such a limit in degenerate or nearly–degenerate semiconductors have been pointed out in the past [10], [11], although only a moderate research effort has walked this alley. Very recently, Ikeda and coworkers [12], [13] disclosed an unexpected increase of the Seebeck coefficient in P-doped silicon-on-insulator (SOI) structures occurring at electron densities $n$ in excess of $3.5 \times 10^{19}$ $cm^{-3}$, claiming for $\alpha(n)$ a maximum at $n \approx 7 \times 10^{19}$ $cm^{-3}$. Authors proposed a semi-quantitative model for the phenomenon that correlates the variation of the Seebeck coefficient with the slope of the density of states $g(E)$ at the Fermi energy $E_F$.

In this paper we report the results of an investigation of the electronic transport properties of degenerate, B-doped polycrystalline silicon structures, confirming the occurrence of a regime where both the Seebeck coefficient and the conductivity increase. This finding possibly acquires a relevance because the possibility of using poly layers instead of single crystal-based SOI would result in prospective devices cheaper by at least two orders of magnitude. A preliminary analysis of both transport coefficients shows that, in order to quantitatively account for the trends observed, a model encompassing both the formation of an impurity band and of a valence band tail must be advanced. Therefore, account must be preliminarily given of the overall physical chemistry of the system.

## II. EXPERIMENTAL PROCEDURE

Films of polycrystalline silicon (450 nm thick, $50 \times 5$ mm$^2$) deposited onto oxidized Si substrates were implanted with boron through an Al sacrificial layer with a fluence of $2 \times 10^{16}$ cm$^{-2}$ at an energy of 60 keV, and were then annealed at 1050 °C for 30 s. This led to a total boron density of $4.4 \times 10^{20}$ cm$^{-3}$. The samples were characterized by resistivity and Seebeck coefficient measurements. To this aim, Al pads were evaporated onto the poly layer. Resistivity was determined by current-voltage characteristics at 20 °C. Seebeck voltages were measured using the integral method [14] by fixing the temperature of the cold contact at 0 °C while heating the other contact between 40 and 120 °C. Each set of Seebeck voltage measurements was repeated on the same sample at least three times to ensure data reliability. Also, Seebeck coefficients were measured on nominally identical samples and found to be reproducible within ±1%.

After removal of the Al sacrificial layer, polycrystalline samples were submitted to thermal treatments. A sequence of annealing cycles in Ar (Experiment I) was carried out at temperatures $T_a$ from 500 to 1000 °C in 50 °C steps, each treatment lasting one hour. After each annealing Al pads were evaporated through a shadow mask and the sample was submitted to the electrical characterization. Then, contacts were removed by HCl etch prior to the subsequent annealing.

In order to check the consistency of the models developed (vide infra), an additional set of samples was prepared and used for Hall measurements. They were obtained by cutting square $17 \times 17$ mm$^2$ chips and evaporating aluminum contacts on small areas in the four corners according to the van der Pauw geometry. Chips were processed in the same way as the samples of Experiment I, except the annealing was carried out in 100 °C steps lasting two hours (hereafter Experiment II). Hall measurements were carried out at room temperature with a maximum magnetic field of 0.5 T.

## III. RESULTS AND DISCUSSION

### A. Overview of the Experimental Results

As anticipated, a first inspection of the experimental data shows some unexpected features. First, should the material obey Ioffe equation, one would expect the Seebeck coefficient to decrease with conductivity. In diluted p-type semiconductors thermopower depends upon carrier density as

$$\alpha = \frac{k_B}{e} \left( \frac{E_V}{k_B T} - A + \ln\left( \frac{2(2\pi m_h^* k_B T)^{3/2}}{h^3 p} \right) \right) \quad (1)$$

where $A$ is a suitable constant depending on the carrier scattering mechanism, $E_V$ is the energy of the top of the valence band, $m_h^*$ is the hole effective mass, $T$ is the absolute temperature, $-e$ is the electron charge, $k_B$ and $h$ are the Boltzmann and Planck constants. Taking the conductivity proportional to the carrier density, namely disregarding the dependence of the drift mobility $\mu$ on $p$ [15] one can write

$$\alpha = \frac{k_B}{e} \left( \frac{E_V}{k_B T} - A + \ln\left( \frac{2e\mu(2\pi m_h^* k_B T)^{3/2}}{h^3 \sigma} \right) \right) \quad (2)$$

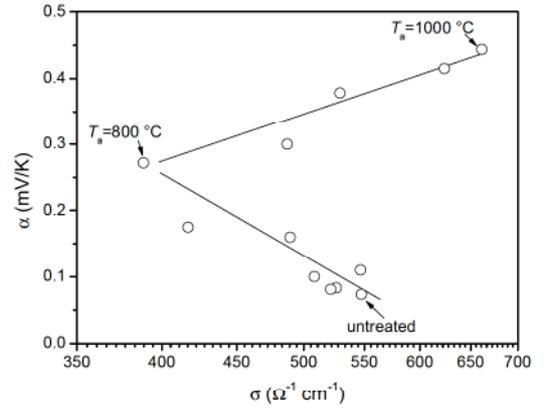

Fig. 1. Experimental variation of the Seebeck coefficient with the conductivity. For annealing temperatures in excess of 800 °C both transport coefficient concurrently grow up.

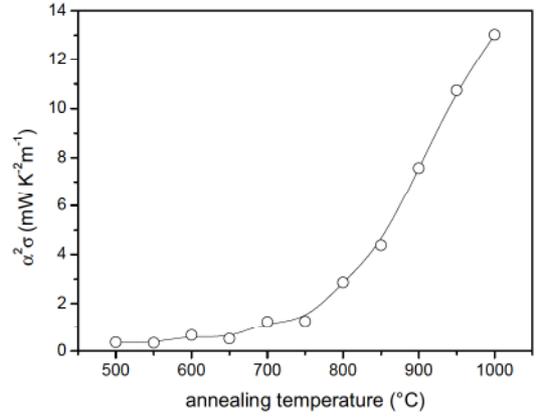

Fig. 2. Power factor of the poly layer as a function of the annealing temperature. The simultaneous growth of $\alpha$ and $\sigma$ lead to a maximum power factor of 13 mW K$^{-2}$ m$^{-1}$. For the sake of comparison, Si nanowires are reported to have power factors of 3.2 mW K$^{-2}$ m$^{-1}$ [1].

Actually, the experimental data displays (Fig. 1) the expected decrease of $\alpha$ with $\sigma$ only up to $T_a$ = 800 °C, the trend then reverting to a simultaneous increase of both conductivity and thermopower — a feature leading to a power factor of 13 mW K$^{-2}$ m$^{-1}$ for $T_a$ = 1000 °C (Fig. 2).

The striking inconsistency of the experimental data with the standard theory of Seebeck effect quite obviously indicates that in degenerate poly-Si the (naive) extension of Ioffe theory, which is appropriate for diluted semiconductors, is of no use, a more detailed analysis of the transport coefficients being required. On the other side, the fortunate concurrent growth of $\alpha$ and $\sigma$ suggests that degenerate Si may demonstrate to be a possibly underestimated candidate for high efficiency thermoelectric generation.

In the next Subsections a model of transport will be outlined, moving from an analysis of the actual concentration of electrically active dopant.

### B. The physical chemistry of the poly layer

In the approach implemented in this work, the thermal history of the samples sets the actual values of carrier density. Thus, any analysis of the transport coefficients requires a preliminary evaluation of the $B_{Si}$ density evolution upon annealing.

Upon implantation and following post-implantation rapid thermal annealing (RTA), boron in excess to its solubility limit at the RTA temperature precipitates. Hall measurements (Experiment II) report actually a carrier density after sequential heat treatment up to 1000 °C of $5.6 \times 10^{19}$ cm$^{-3}$, smaller than the solubility threshold of single crystals ($1.2 \times 10^{20}$ cm$^{-3}$). Actual solubility of substitutional impurities in polycrystalline semiconductors may eventually exceed the corresponding value for single crystals because of the combined effect of lattice strain [16] and of the destabilization of segregates [17]. Instead, in the absence of significant inner surface (GB) oxidation, it is more difficult to justify a decrease of solubility in poly-Si. It is therefore sensible to ascribe such a difference to compensation. The donor density $c_D$ can be computed as the difference between the $B_{Si}$ single-crystal solid solubility at 1000 °C and the actual carrier density upon the final annealing treatment, leading to $c_D = 6.3 \times 10^{19}$ cm$^{-3}$. At this stage no information is available about the chemical nature of donor impurities.

As boron concentration exceeds the solid solubility at any annealing temperature, thermal processing promotes its precipitation. Modeling of the sequential isochronous annealing cycles can be carried out within the frame of the Ham's theory [18]. Accounting for the growth of precipitates with time in the spherical particle approximation, the average concentration of the dopant at time $t$, $\bar{c}(t)$, computes to

$$H\left(\left[\frac{\bar{c}(0)-\bar{c}(t)}{\bar{c}(0)-c_s}\right]^{\frac{1}{3}}\right) = \frac{Dt}{r_s^2}\left(\frac{\bar{c}(0)-c_s}{c_p-c_s}\right)^{\frac{1}{3}} \quad (3)$$

where the Ham function is defined as

$$H(u) := \frac{1}{6}\ln\left(\frac{u^2+u+1}{u^2-2u+1}\right) - \frac{1}{\sqrt{3}}\tan^{-1}\left(\frac{2u+1}{\sqrt{3}}\right) + \frac{1}{\sqrt{3}}\tan^{-1}\left(\frac{1}{\sqrt{3}}\right), \quad (4)$$

$c_s$ is the solubility of the dopant, $c_p$ is its concentration in the precipitate ($8\times 10^{22}$ cm$^{-3}$ for SiB$_4$), $r_s$ is the average spacing of precipitates (correlated to their number per volume unit $N_p$ as $r_s = [(4/3)\pi N_p]^{-1/3}$), and $D$ is the dopant diffusivity. Using the standard theory of nucleation, $N_p$ can be correlated with $c$ as

$$N_p = K_p \exp\left[-\frac{16\pi}{3}\left(\frac{\gamma}{k_B T}\right)^3 \frac{1}{c_p^2[\ln(c/c_s)]^2}\right] \quad (5)$$

where $K_p$ is proportional to the number of nucleation sites and $\gamma$ is the precipitate surface free energy. For Si:B Solmi et al. [19] proposed values for $K_p$ and $\gamma$ of $10^{16}$ cm$^{-3}$ and $2\times 10^{-5}$ J cm$^{-2}$, resp.. One estimates $N_p(T)$ to range in turn between $5.8 \times 10^{15}$ cm$^{-3}$ (at 500 °C) and $1.0 \times 10^{15}$ cm$^{-3}$ (at 1000 °C). In the sequential processing adopted in this work it is therefore sensible to disregard any nucleation event further to that induced by the first annealing at 500 °C, upon which nuclei distribute in the solid solution with an average distance $r_s$ of 35 nm. Their critical radius computes to

$$r_{cr} = \frac{2\gamma}{c_p(k_B T)\ln(c/c_s)}, \quad (6)$$

i.e. $r_{cr} \approx 0.9$ nm for $c \approx 10^{20}$ cm$^{-3}$. Solving Eq. (3) for $\bar{c}(t)$ and in view of the fact that each annealing step lasted the same time $t_a$, the average concentration $c_j$ of dissolved (non precipitated) boron at the $j$-th processing step can be written as

$$\left(\frac{c_j-c_{j-1}}{c_s-c_{j-1}}\right)^{\frac{1}{3}} = H^{-1}\left(\frac{t_a D_j}{r_s^2}\left[\frac{c_{j-1}-c_s}{c_p-c_s}\right]^{\frac{1}{3}}\right) \quad (7)$$

where $D_j$ is short for $D(T_j)$.

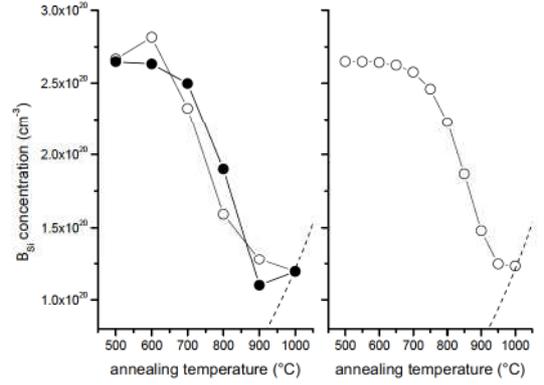

Fig. 3. Plot of *(left)* the experimental (open circles) and fitted (filled circles) $B_{Si}$ concentration upon annealing (Experiment II); and *(right)* of the computed $B_{Si}$ density as a function of the annealing temperature (Experiment I). The dashed line shows the solubility of $B_{Si}$ in the single crystal [20], [17]. Full lines serve as eye's guide only.

Eq. (7) was fitted to data from Expriment II under the assumption $p = c - c_D$ (Fig. 3, left). Boron diffusivity is found to be

$$D_B = (4.8 \times 10^{-10} \text{cm}^2\text{s}^{-1})\exp\left(-\frac{1.11\text{eV}}{k_B T}\right)$$

The relatively low value of the activation energy may find a rationale in the presence of preferential diffusion paths for boron at GBs. Based upon these results, we could estimate also the $B_{Si}$ evolution in Experiment I (Fig. 3, right). Also for this set, annealing is found to promote diffusion-limited precipitation all over the explored range of temperatures.

### C. Transport coefficients in degenerate silicon

As for the analysis of the boron content in the film, samples from Experiment II were used to support the development of a model for hole transport in the poly layer.

The temperature-coefficient of the hole density (not shown) confirms that the poly film is degenerate in the whole boron density range accessed by the annealing cycles. Following Fritzsche [21] and neglecting electron correlation effects one can write

$$\sigma = e \int_{-\infty}^{+\infty} g(E)\mu(E)f(E,E_f)[1 - f(E,E_f)]dE \quad (8)$$

where $g(E)$ is the density of states (DOS), $\mu(E)$ is the (energy-resolved) carrier mobility, and $f(E,E_f)$ is the Fermi-Dirac function. Likewise, the Seebeck coefficient can be written as

$$\alpha = -\frac{k_B}{e} \int_{-\infty}^{+\infty} \frac{E - E_f}{k_B T} \frac{\sigma(E)}{\sigma} dE \quad (9)$$

where $\sigma(E)$ is the integrand in the right-hand side of Eq. (8). In metallic conductors it is customary to assume that only the states within $k_B T$ from $E_f$ contribute to charge transport. Thus, Eq. (8) simplifies to

$$\sigma = e(k_B T) g(E_f) \mu(E_f) \quad (10)$$

while, expanding $g(E)\mu(E)$ around $E_f$, one gets

$$\alpha = \frac{\pi^2}{3} \frac{k_B}{e} (k_B T) \left( \frac{d \ln(g(E)\mu(E))}{dE} \right)_{E=E_f} \quad (11)$$

which is the celebrated Mott's formula for thermopower in metallic conductors.

Since both Eqs. (10) and (11) (as well as Eqs. (8)–(9)) directly associate $\sigma$ and $\alpha$ to the system electronic structure, they allow to evaluate how transport coefficients get affected by the modification of the band structure due to heavy doping.

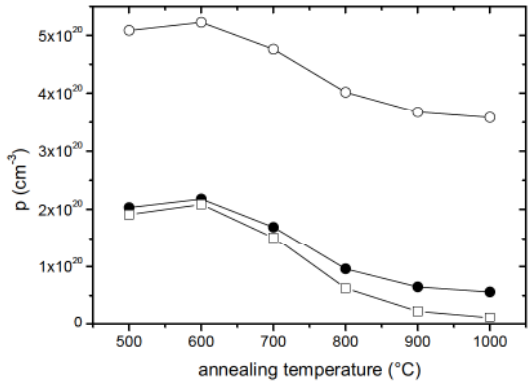

Fig. 4. Experimental evolution (filled circles) of the carrier density upon annealing (Experiment II) compared to hole density computed according to the standard Mott approximation (open circles) and accounting for the exact carrier statistics (open squares). See text for details. Lines are intended as eye's guide only.

Modeling of the band structure of heavily doped semiconductors is still at present a complex and challenging task. In strongly degenerate p-type semiconductors the most significant effect deriving from the interaction of the dopant with the crystal atoms is band gap narrowing (BGN) [22]. Without any pretension of completeness or cogency, of the several factors contributing to BGN we will consider here only the Coulomb interaction potential $V$ between ionized acceptors and holes. The average interaction potential shifts the valence band edge by

$$\Delta E = \frac{2e^2 \exp(-d/2\lambda)}{4\pi\epsilon(d/2)} - \frac{e^2 \exp(-d/\lambda)}{4\pi\epsilon d} \quad (12)$$

where $\epsilon$ is the static permittivity of the medium, $d$ is the acceptor spacing, and $\lambda$ is the screening length. Since ionized impurities are randomly distributed through the solid, they also induce potential fluctuations which are responsible for an additional band edge tailing. If the potential distribution $q(V)$ is Gaussian with a standard deviation

$$\delta V = \left( \frac{c_D + c}{8\pi^2 \epsilon^2} e^4 \lambda \right)^{1/2} \quad (13)$$

then the local DOS at a point of potential $V$

$$g_V^{(0)}(E) = \frac{24\sqrt{2} m_h^{3/2}}{h^2} (V - E)^{1/2} \quad (14)$$

gets convoluted with $q(V)$ as

$$g_V(E) = \int_E^{+\infty} g_V^{(0)}(E) q(V) dV \quad (15)$$

Finally, interactions among impurities set up an impurity band, contributing to the DOS a term of the form [22]

$$g_A(E) = \frac{c}{B} \int_{E-B/2}^{E+B/2} q(V) dV \quad (16)$$

where

$$B = 2 \left| \int_0^\infty J(R) 4\pi c R^2 \exp(-4\pi c R^3/3) dR \right| \quad (17)$$

with

$$J(R) = \frac{e^2}{4\pi\epsilon} (1 + \xi R) \exp(-\xi R) \quad (18)$$

and $\xi^{-1} = a_H(E_0/E_B)$ is the generalized Bohr radius of boron, $a_H$ is the hydrogen Bohr radius, $E_0 = c^2/(8\epsilon a_H)$, and $E_B$ is the energy difference between the boron level and the edge of the valence band in the low doping limit ($E_B = 45$ meV).

*1) Carrier density:* The position of the Fermi level can be easily obtained by imposing charge conservation:

$$\int_{-\infty}^{+\infty} g_V(E) f(E,E_f) dE = \int_{-\infty}^{+\infty} [g_V(E) + g_A(E)] dE \quad (19)$$

Once the dependence of both the DOS and the Fermi level upon boron density is known, it is immediate to compute all transport coefficients. We actually used the data from Experiment II to verify the accuracy of the DOS model. As a first check of consistency, Fig. 4 compares the experimental carrier density (as obtained by Hall measurements) with both the carrier density computed in the standard Mott approximation, $g(E_f)(k_B T)$, and that obtained accounting for the full contributions of all states, $\int_{-\infty}^{+\infty} g(E)[1 - f(E,E_f)]dE$. While both computations catch the evolution of $p$ with annealing, only the latter quantitatively recovers the experimental trend. Note that both models are neither parametrized or anyway fitted to data, what makes the agreement especially significant.

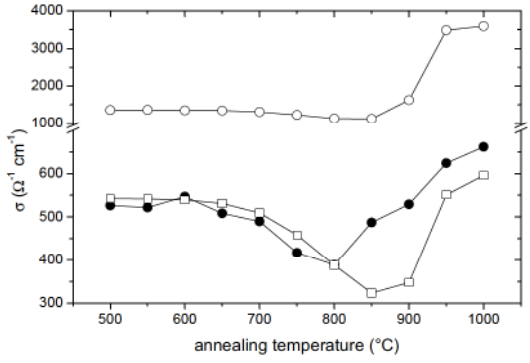

Fig. 5. Comparison of the experimental (filled circles) conductivity data from Experiment I with simulations according to Eq. (10) (open circles) and Eq. (8) (open squares). In both cases simulations are based upon the computed $B_{Si}$ concentration profiles obtained from Eq. (7). Mobility was estimated by interpolating experimental mobility data from Experiment II over $p$ at the pertinent approximation level.

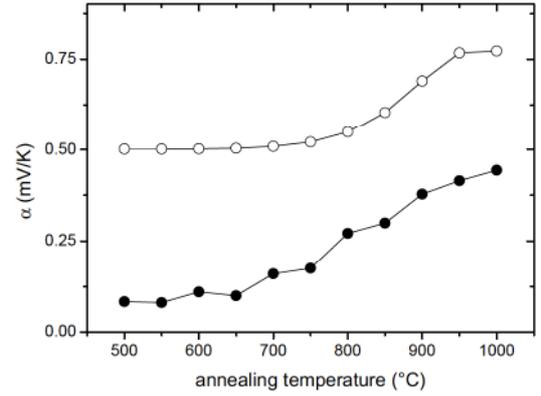

Fig. 6. Computed (open circles) and experimental (filled circles) Seebeck coefficients upon annealing (Experiment I). Lines are eye's guides.

At this level of confidence, boron concentrations obtained in Sect. III-B from the analysis of diffusion-limited precipitation on Experiment I can also be used to predict in turn the carrier density as a function of $T_a$, getting a steadily decrease of the computed $p$ upon subsequent annealing steps.

*2) Mobility and conductivity:* A full analysis of mobility is beyond the scope of this paper. An evaluation of $\mu(E)$ is actually made very complicated by the simultaneous presence of GB and by the presence of sparse precipitates in the layer. While we set to postpone an analysis of this phenomenon, it may suffice here to say that drift mobility (as obtained by Hall measurements in Experiment II) is limited by GB scattering at low $T_a$, recovering values close to that of the single crystal for $T_a > 800$ °C. Interpolating the experimental mobility from Experiment II over the computed carrier density and in view of the computed $p$ values for Experiment I we can obtain the expected trend of $\sigma(T_a)$ in both the standard approximation [Eq. (10)] and accounting for the whole carrier contribution [Eq. (8)] — comparing it to the experimental conductivity (Fig. 5). Differently from the the carrier density case, we can qualitatively reproduce the evolution of the experimental data only using the full expression of $\sigma$.

*3) Thermoelectric power:* Finally, thermopower dependence upon carrier density follows from Eq. (11). Evolution of $[\mathrm{d}\ln g(E)/\mathrm{d}E]_{E=E_f}$ is trivial. Instead, the mobility term $[\mathrm{d}\ln \mu(E)/\mathrm{d}E]_{E=E_f}$ can be here just estimated considering that only the functional dependence of $\mu$ on $E$ survives the log-derivative. Thus, since impurity scattering sets $\mu$ proportional to $E^{3/2}$,

$$\left(\frac{\mathrm{d}\ln\mu(E)}{\mathrm{d}E}\right)_{E=E_f} = \frac{3/2}{E_f} \qquad (20)$$

We found that $\alpha$ is properly predicted to increase with $T_a$, although the computation only recovers such a trend in a qualitative way (Fig. 6). In this case, a computation of $\alpha$ using Eq. (9) cannot be put forward, since it requires a full model of $\mu(E)$ that is not currently available for poly Si. It seems sound to speculate that, as in the previous two cases, Mott equation is qualitatively (but not quantitatively) adequate in depicting the transport processes in degenerate (polycrystalline) silicon.

It might be worth noting that the observed enhancement of the Seebeck coefficient is a perhaps counterintuitive effect of the high doping level. The modification of $g(E)$ around the Fermi energy caused by band tailing actually results in an increase of the $\mathrm{d}\ln g(E)/\mathrm{d}E$ term in the Mott equation compared to more lightly doped Si.

## IV. SUMMARY AND CONCLUDING REMARKS

We reported a first analysis of the transport properties of highly doped p-type polycrystalline silicon films. The material shows a somehow unexpected concurrent increase of both its conductivity and Seebeck coefficient upon annealing at temperatures higher than 800 °C. This leads to very interesting power factors. Since GBs are expected to lower the thermal conductivity $\kappa$, should we recover the values reported in the literature for heavily doped polycrystalline Si of $\kappa = 0.34$ W cm$^{-1}$ K$^{-1}$ [23], the material would display a $zT$ value of 0.11 at 300 K, a figure of some relevance in consideration of the low material cost.

Carrier density, $\sigma$ and $\alpha$ could be simulated using a band model accounting for BGN and band tailing. Quantitative agreement with experimental data could be obtained only relaxing Mott approximation. This can be explained considering the (relatively) low carrier density of degenerate Si in comparison to standard metals. Additional support to the analysis of the system will require an accurate modeling of the mobility, which is complicated by the presence of GB and, even more, by the occurrence of second phase precipitation in the layer. Preliminary analyses seem to suggest that polycrystallinity might favorably modify the functional form of $\mu(E)$. Further characterization of the layer is currently under way and will be reported in forthcoming papers.

### ACKNOWLEDGMENTS

The authors are indebted with Dr. Maurizio Acciarri for his support in the setup of Hall effect measurements and with Prof. Stefano Frabboni for preliminary TEM analyses.

## REFERENCES

[1] A. I. Hochbaum, R. K. Chen, R. D. Delgado, W. J. Liang, E. C. Garnett, M. Najarian, A. Majumdar, and P. D. Yang, "Enhanced thermoelectric performance of rough silicon nanowires," *Nature*, vol. 451, no. 7175, pp. 163–167, 2008.

[2] A. I. Boukai, Y. Bunimovich, J. Tahir-Kheli, J. K. Yu, W. A. Goddard, and J. R. Heath, "Silicon nanowires as efficient thermoelectric materials," *Nature*, vol. 451, no. 7175, pp. 168–171, 2008.

[3] M. G. Kanatzidis, "Nanostructured thermoelectrics: The new paradigm?" *Chem. Mater.*, vol. 22, no. 3, pp. 648–659, 2009.

[4] A. Ioffe, *Semiconductor Thermoelements and Thermoelectric Cooling*. London: Infosearch Ltd., 1957.

[5] T. H. Geballe and G. W. Hull, "Seebeck effect in silicon," *Phys. Rev.*, vol. 98, no. 4, p. 940, 1955.

[6] W. Fulkerson, J. P. Moore, R. K. Williams, R. S. Graves, and D. L. McElroy, "Thermal conductivity, electrical resistivity, and Seebeck coefficient of silicon from 100 to 1300 °K," *Phys. Rev.*, vol. 167, no. 3, p. 765, 1968.

[7] G. A. Slack and M. A. Hussain, "The maximum possible conversion efficiency of silicon-germanium thermoelectric generators," *J. Appl. Phys.*, vol. 70, no. 5, pp. 2694–2718, 1991.

[8] G. D. Mahan, "Good thermoelectrics," in *Solid State Physics*, ser. Solid State Physics-Advances in Research and Applications. San Diego: Academic Press Inc, 1998, vol. 51, pp. 81–157.

[9] G. F. Cerofolini, M. Ferri, E. Romano, A. Roncaglia, E. Selezneva, A. Arcari, F. Suriano, G. P. Veronese, S. Solmi, and D. Narducci, "Industrially scalable process for silicon nanowires for Seebeck generation," 2010, presented at this Conference.

[10] L. Weber and E. Gmelin, "Transport properties of silicon," *Appl. Phys. A*, vol. 53, no. 2, pp. 136–140, 1991.

[11] M. E. Brinson and W. Dunstant, "Thermal conductivity and thermoelectric power of heavily doped n-type silicon," *J. Phys. C*, vol. 3, no. 3, p. 483, 1970.

[12] F. Salleh, K. Asai, A. Ishida, and H. Ikeda, "Seebeck coefficient of ultrathin silicon-on-insulator layers," *Appl. Phys. Express*, vol. 2, p. 071203, 2009.

[13] H. Ikeda and F. Salleh, "Influence of heavy doping on Seebeck coefficient in silicon-on-insulator," *Appl. Phys. Lett.*, vol. 96, no. 1, pp. 012 106–3, 2010.

[14] C. Wood, A. Chmielewski, and D. Zoltan, "Measurement of Seebeck coefficient using a large thermal-gradient," *Rev. Sci. Instrum.*, vol. 59, no. 6, pp. 951–954, 1988.

[15] Considering the dependence of $\mu$ on $p$, e.g. according to the Brooks-Herring model [24], leads to even more paradoxical results. Since $\mu \propto p^{-1}$ one obtains that $\alpha$ is independent of $\sigma$.

[16] J. Adey and et al., "Enhanced dopant solubility in strained silicon," *J. Phys.: Condens. Matter*, vol. 16, no. 50, p. 9117, 2004.

[17] X. Luo, S. B. Zhang, and S. H. Wei, "Understanding ultrahigh doping: The case of boron in silicon." *Phys. Rev. Lett.*. vol. 90. no. 2. p. 026103. 2003.

[18] F. S. Ham, "Theory of diffusion-limited precipitation," *J. Phys. Chem. Solids*, vol. 6, no. 4, pp. 335–351, 1958.

[19] S. Solmi, E. Landi, and F. Baruffaldi, "High-concentration boron-diffusion in silicon - simulation of the precipitation phenomena," *J. Appl. Phys.*, vol. 68, no. 7, pp. 3250–3258, 1990.

[20] G. L. Vick and K. M. Whittle, "Solid solubility and diffusion coefficients of boron in silicon," *J. Electrochem. Soc.*, vol. 116, no. 8, pp. 1142–1144, 1969.

[21] H. Fritzsche, "A general expression for the thermoelectric power," *Solid State Commun.*, vol. 9, no. 21, pp. 1813–1815, 1971.

[22] T. F. Lee and T. C. McGill, "Variation of impurity- to- band activation energies with impurity density," *J. Appl. Phys.*, vol. 46, p. 373, 1975.

[23] Y. C. Tai, C. H. Mastrangelo, and R. S. Muller, "Thermal conductivity of heavily doped low-pressure chemical vapor deposited polycrystalline silicon films," *J. Appl. Phys.*, vol. 63, no. 5, pp. 1442–1447, 1988.

[24] D. Chattopadhyay and H. J. Queisser, "Electron scattering by ionized impurities in semiconductors," *Rev. Mod. Phys.*, vol. 53, no. 4, p. 745, 1981.